\documentclass[english,11pt,aps,prd,a4paper,preprintnumbers,floatfix,nofootinbib,showpacs,superscriptaddress, notitlepage]{revtex4-1} 

\newcommand{\AddrUNAM}{Instituto de F\'isica, Universidad Nacional Aut\'onoma de M\'exico, A.P. 20-364, Ciudad de M\'exico 01000, M\'exico.}
\newcommand{\AddrINFN}{Istituto Nazionale di Fisica Nucleare, Sezione di Bari, Via Orabona 4, 70126 Bari, Italy}

\newcommand{\AddrTEC}{Tecnol\'ogico Nacional de M\'exico/ITS de Jerez, C.P. 99863, Zacatecas, M\'exico.}

\usepackage{booktabs}
\usepackage{color}
\usepackage{graphicx}
\usepackage{multirow, tabu}
\usepackage{amsmath}
\usepackage{amssymb}
\usepackage[colorlinks=true,citecolor=darkred,urlcolor=darkred, pdfborder={0 0 0}]{hyperref}
\usepackage[normalem]{ulem}
\definecolor{darkred}{rgb}{0.6,0,0}
\definecolor{drkgrn}{RGB}{0, 51, 0}
\definecolor{gray}{RGB}{128, 128, 128}
\usepackage{soul}
\def\beq{\begin{equation}}
\def\eeq{\end{equation}}

\newcommand{\projl}{(1 - \gamma^5)}
\newcommand{\projr}{(1 + \gamma^5)}

\begin{document}

\title{CE$\nu$NS as a probe of flavored generalized neutrino interactions}

\author{L.  J.  Flores}\email{ljflores@jerez.tecnm.mx}\affiliation{\AddrTEC}
\author{Newton Nath}\email{newton.nath@ba.infn.it}\affiliation{\AddrINFN}\affiliation{\AddrUNAM}
\author{Eduardo Peinado} \email{epeinado@fisica.unam.mx}\affiliation{\AddrUNAM}

\begin{abstract}
{\noindent
We examine the potential to probe  {\it generalized neutrino interactions} (GNI),  exotic effective couplings  due to  new physics interactions beyond the Standard Model, in the coherent-elastic neutrino-nucleus scattering experiments in light of the latest COHERENT-CsI,  and -LAr data.  Our analysis focuses on scalar, vector and tensor flavored-GNI parameters.  A combined  analysis has been made to constrain these exotic couplings for the CsI and LAr detector. We further add  the projected forthcoming reactor-based  Scintillating Bubble Chamber detector to examine these couplings. 
It has been observed that the addition of reactor data strongly constrained electron flavor GNI.
}
\end{abstract}

\maketitle

%%%%%%%%%%%%%%%%%%%%%%%%%%%%%%%%%%%%%%%
\section{Introduction}
%%%%%%%%%%%%%%%%%%%%%%%%%%%%%%%%%%%%%%%
 With the confirmation that neutrinos possess non-zero masses and their different flavors are mixed  from different phenomenal neutrino oscillations experiments,  neutrino physics  has entered in  the era of precision measurements~\cite{Zyla:2020zbs}.     
Given the capability and sensitivity reach of various ongoing as well as forthcoming neutrino experiments, 
a natural question arises, are there any new physics that could significantly impact the standard neutrino interactions?
Among various new physics scenarios that appear in the form of physics beyond the Standard Model (SM), the neutrino non-standard interactions (NSI) have garnered lots of attention in recent times (see \cite{Ohlsson:2012kf, Miranda:2015dra, Farzan:2017xzy} for recent reviews). They are traditionally described by the dimension six effective operators and have the (chiral) vector form,  originally proposed in \cite{Wolfenstein:1977ue}.  

Moreover, possibilities of having more {\it exotic interactions} involving neutrinos such as the scalar,  pseudoscalar,  axial-vector, and tensor Lorentz structures can not be ignored \cite{Bergmann:1999rz}.   These exotic interactions are commonly referred as {\it generalized neutrino interactions} (GNI)~\cite{Lindner:2016wff,Rodejohann:2017vup,Bischer:2018zcz}.
Many attempts have been made in recent times to explore these interactions in neutrino oscillations \cite{Bischer:2018zcz, Khan:2019jvr}, and neutrino scattering \cite{Rodejohann:2017vup, Kosmas:2017tsq,AristizabalSierra:2018eqm,Chen:2021uuw,Li:2020lba} experiments.  For the latest global constraints of GNI parameters see Ref.~\cite{Escrihuela:2021mud}. 
Here, we are interested to investigate the impact of these exotic interactions in the coherent-elastic neutrino-nucleus scattering (CE$\nu$NS) experiments. These modify the SM effective couplings and eventually change the measurements of the SM weak-interaction differential cross-section.
 
The CE$\nu$NS, an allowed SM process, was originally proposed in the early seventies~\cite{Freedman:1973yd,Freedman:1977xn}, and it was first detected in 2017 by the COHERENT collaboration using  cesium-iodide (CsI) detector~\cite{Akimov:2017ade}  at $6.7^{}\sigma$ significance. 
They recently published their latest measurements using the same detector~\cite{Akimov:2021dab} and reported $306\pm20$   CE$\nu$NS events,  which  are consistent  with the SM predictions of $341 \pm 11 ({\rm theory}) \pm 42 ({\rm expt.})$ events at $1^{}\sigma$.
The COHERENT collaboration has also reported the first measurement of CE$\nu$NS process using CENNS-10 liquid argon detector~\cite{Akimov:2020pdx,Akimov:2020czh}  with a significance greater than $3^{}\sigma$. 
In the SM, the CE$\nu$NS process is very well understood which is induced by the exchange of $Z$ boson~\cite{Freedman:1973yd}.
The thumb rule to occurs such scattering is that the product of three-momentum transfer ($|\overrightarrow{q}| $) and nuclear radius ($ R $) should satisfy $(|\overrightarrow{q}| R \ll 1 )$,  which needs neutrino energies below 50 MeV.  Hence,  such scattering experiments face major challenges   to observe  nuclear recoils with a very small kinetic energy ($ \sim $ keV).
It is very important to analyze this process because of its ability to probe the SM parameters at low momentum transfer~\cite{Scholberg:2005qs,Lindner:2016wff,Deniz:2017zok,Miranda:2019wdy},  
nuclear physics parameters~\cite{Cadeddu:2017etk,Papoulias:2019lfi,Canas:2019fjw,Cadeddu:2020lky},  neutrino  electromagnetic properties~\cite{Cadeddu:2018dux,Miranda:2019wdy},  sterile neutrinos~\cite{Miranda:2019skf,Blanco:2019vyp,Berryman:2019nvr,Miranda:2020syh}, neutrino non-standard interactions~\cite{Liao:2017uzy,Denton:2018xmq,Giunti:2019xpr,Esteban:2019lfo,Coloma:2019mbs,Khan:2019cvi, Flores:2020lji,Denton:2020hop,Du:2021rdg} as well as sub-GeV dark matter models \cite{COHERENT:2019kwz,delaVega:2020xcu,Majumdar:2021vdw}.

In this work, motivated by the latest COHERENT measurements a combined analyses of the CsI, and LAr data are made in order to examine GNI parameters. 
We perform a detailed numerical analysis  to constrain these parameters in a model-independent way, which allows one to study various new physics models that can be tested or rejected in the near future. 
In doing so, we first calculate the SM events for both  detectors and later calculate the same in presence of GNI parameters.  We then compare these events with the observed data. 
In order to investigate the parameter space of these GNI, we perform a chi-square analysis considering one non-zero GNI at a time. 
Here, we aim to examine flavorful scalar, vector and tensor GNI, i.e. for electron and muon flavors \footnote{Notice that in our analysis we do not distinguish neutrinos from anti-neutrinos.}.
This has been done for both detectors as well as their combined analysis.  Later, we proceed to perform numerical analysis for two non-zero GNI at a time, which are presented in contours at 1, 2, and 3$^{} \sigma $ significance level.  
In addition, we also explore the potential of forthcoming reactor-based CE$\nu$NS experiment  from the Scintillating Bubble Chamber (SBC) Collaboration \cite{Flores:2021ihw}.
  A combined analysis  of SBC with COHERENT data helps us to constrain these exotic effective couplings in a comprehensive manner. 

We organize this work as follows.  Sec.  \ref{sec:GNI} is dedicated to the general description of generalized neutrino interactions.  Events number calculation for the neutrino-nucleus scattering process is discussed in Sec. \ref{sec:coherent}, where we first calculate number of events in the SM and subsequently present our results including GNI.  Detailed discussions about the chi-square analysis are presented in Sec.  \ref{sec:Nu-Analysis}.   
Summary of this work and concluding remarks are presented in Sec.  \ref{sec:Conclusion}. In appendices \ref{app:GNI-Param}, and \ref{app:LAr_analyses} we briefly discuss about the parametrization of GNI parameters as well as different approaches of $ \chi^{2} $  analysis for the LAr detector, respectively.
%%%%%
\section{Generalized Neutrino Interactions}\label{sec:GNI}
%%%%%%%%%%%%%%%%%%%%%%%%%%%%%%%%%%%%%%%%%%%%%%%%%%%%%%%%%%%%%%%%%%%%
We start our discussion by summarizing the elastic neutrino scattering in the SM, where neutrinos undergo both the neutral current (NC) and charged current (CC) interactions.  As the standard matter composed of electron,  hence $ \nu_e $ possesses both the  NC and CC interactions, whereas $ \nu_{\mu} $  and $ \nu_{\tau} $ undergo only NC interactions. 
The effective SM Lagrangian for the NC interactions can be formulated as 
%%%%%%%%%%%%%%
\begin{equation}\label{eq:SM-NC-Inter}
  \mathcal{L}^{\rm NC}_\text{SM} \supset  \dfrac{G_F}{\sqrt{2}} \left[ \overline{\nu_{\alpha}} \gamma^{\rho} (1 - \gamma^{5}) \nu_{\alpha} \right] 
        \left[  \bar{f} \gamma_{\rho}  (g^{f}_{V}  - g^{f}_{A} \gamma^{5}) f \right] 
  + \text{h.c.} \;,
\end{equation}
%%%%%%%%%%%%%
where $  \nu_{\alpha} = \nu_e,  \nu_{\mu} $ or $ \nu_{\tau} $ flavor,   $ f =  e,  p$  or $ n $,  and $G_F$ is the Fermi constant.  Also,  the vector ($ g^{f}_{V} $) and axial vector ($ g^{f}_{A} $) couplings are given by
%%%%%%%%%
\begin{align}\label{eq:SM-Couplings}
g^{e}_{V} &  = - \dfrac{1}{2} + 2 \sin^{2} \vartheta_{w},   ~  ~  g^{p}_{V}  =  \dfrac{1}{2} - 2 \sin^{2} \vartheta_{w},   ~  ~ g^{n}_{V}  =  - \dfrac{1}{2}  \nonumber \;, \\
g^{e}_{A} &  = - \dfrac{1}{2},   ~  ~  g^{p}_{A}  =   \dfrac{1}{2} ,  ~  ~  g^{n}_{A}  =  - \dfrac{1}{2} \;,
\end{align}
%%%%%%%%%%%
where $ \vartheta_{w} $ is the Weinberg angle.   
Similarly,   for the CC interactions, we can express $ \nu_e - e$ scattering effective Lagrangian as
\begin{equation}\label{eq:SM-CC-Inter}
  \mathcal{L}^{\rm CC}_\text{SM} \supset  \dfrac{G_F}{\sqrt{2}} \left[ \overline{\nu_{e}} \gamma^{\rho} (1 - \gamma^{5}) \nu_{e} \right] 
        \left[  \bar{e} \gamma_{\rho}  (1 - \gamma^{5}) e \right] 
  + \text{h.c.} \;
\end{equation}

However, an interesting proposal has been made in \cite{Wolfenstein:1977ue} that  interactions of neutrinos with matter can lead to  \textit{new physics} beyond the SM, which may appear in  unknown couplings,  usually referred to as  neutrino non-standard interactions (NSI). 
Such interactions are of (chiral) vector  form  and  expressed in literature as
\begin{equation}\label{eq:NSI}
  \mathcal{L}^{\rm NC/CC}_\text{NSI} \supset  \dfrac{G_F}{\sqrt{2}} \epsilon^{ff^{\prime}  }_{\alpha\beta}  \left[ \overline{\nu}_\alpha \gamma^{\rho} (1 - \gamma^{5}) \nu_\beta \right] 
\left[ \bar{f} \gamma_{\rho}  (1 \mp \gamma^{5}) f^{\prime}\right] 
  + \text{h.c.} \;,
\end{equation}
where $ \epsilon^{ff^{\prime}}_{\alpha\beta}$ are the NSI parameters,  $  f,  f^{\prime} = e, u, d$,  and  $ \alpha, \beta = e, \mu, \tau $.  
Also,   $ f = f^{\prime} $ gives NC NSIs, whereas CC NSIs are obtained for $ f \neq f^{\prime} $. 
Notice that according to the Fermi theory within the SM, we have  $ G_F/\sqrt{2} \simeq g^{2}_W / 8 M^{2}_W$, where $g_W$ and $M_W$ are the  weak coupling constant and $ W $ boson mass,  respectively. 
Drawing an analogy,   after heavy degrees of freedom are integrated out, the effective dimension-6 NSI  coupling can be written as  
\begin{equation}\label{eq:NSI-NewPhysics}
\epsilon \simeq  \dfrac{ \sqrt{2}}{ G_F} \dfrac{ g^{2}_X}{M^{2}_X}  \;, 
\end{equation}
where $ g_X $ and $ M_X $ are the coupling strength and mass of the exchanged particle. 

In this work, we are interested to study the most general Lorentz-invariant interactions beyond the usual vector-like form. 
Such  exotic   neutrino-nucleus  interactions  are  called as generalized neutrino interactions,  and 
can be expressed as  \cite{Lindner:2016wff}
\begin{equation}
{\cal L}\supset   \frac{G_{F}}{\sqrt{2}}   \sum_{a=S,P,V,A,T}\overline{\nu}\,\Gamma^{a}\nu\left[\overline{q}\Gamma^{a}(C_{a}^{(q)}+\overline{D}_{a}^{(q)}i\gamma^{5})q\right],\label{eq:Lagrangian-GNI}
\end{equation}
where $q = u, d$ quarks. Also, 
\begin{equation}
\Gamma^{a}=\{I,i\gamma^{5},\gamma^{\mu},\gamma^{\mu}\gamma^{5},\sigma^{\mu\nu}\} ;\hspace{1cm}    \sigma^{\mu\nu}\equiv\frac{i}{2}[\gamma^{\mu},\gamma^{\nu}].\label{eq:DiracBilinear}
\end{equation}

The effective GNI Lagrangian involves  five types of possible interactions,  namely, scalar ($S$),
pseudoscalar ($P$), vector ($V$), axial-vector ($A$), and tensor
($T$) interactions similar to the  NSI Lagrangian given by Eq.\ (\ref{eq:NSI}). 
Here,   coefficients $C_{a}^{(q)}$ and $\overline{D}_{a}^{(q)}$ are  dimensionless and carry information of GNI parameters (see our discussion below). Without loss of generality one can take them as real,  and also $\overline{D}_{a}^{(q)}$  satisfies
\begin{equation}
D_{a}^{(q)}\equiv\begin{cases}
\overline{D}_{a}^{(q)} & (a=S,\thinspace P,\thinspace T)\\
i\overline{D}_{a}^{(q)} & (a=V,\thinspace A)
\end{cases} \;.
\end{equation}
It is important to point out here that there exists another convention to define  the effective GNI Lagrangian  in literature  (see Ref. \cite{Bischer:2018zcz})  and is equivalent to Eq. (\ref{eq:Lagrangian-GNI}),  which takes the following form, 
%%%%%%
\begin{equation}\label{eq:GNI-operators}
{\cal L} \supset  \frac{G_{F}}{\sqrt{2}}  ~  \sum^{10}_{j=1} \widetilde{\varepsilon}^{~ q,   ~ j}_{\alpha \beta \gamma \delta}(  \overline{\nu}_{\alpha} \mathcal{O}_j \nu_{\beta} )  ~ (  \overline{q}_{\gamma} \mathcal{O}^{\prime}_j q_{\delta} ) \;,
\end{equation}
 where $ \widetilde{\varepsilon}^{~ q,   ~ j}_{\alpha \beta \gamma \delta} $ are the GNI parameters, operators $ \mathcal{O} $, and $ \mathcal{O}^{\prime} $ take 10 different forms  as given by Table \ref{tab:GNI-operators} of appendix \ref{app:GNI-Param}.   For instance, it is to be noted that  the scalar GNI,  $ \varepsilon_{S} $,  is defined when $ \mathcal{O}  = (1 - \gamma^{5})$ and  $ \mathcal{O}^{\prime}  = 1$. 
 %%%%%%

Now in order to show the correlation between GNI parameter $ \varepsilon $ (see Eq.~(\ref{eq:GNI-operators})) with coefficients  $C_{a}^{(q)}$ and $\overline{D}_{a}^{(q)}$ of Eq.~(\ref{eq:Lagrangian-GNI}), we expand   Eq.  (\ref{eq:Lagrangian-GNI}) for the  scalar GNI as follows
 \begin{align} \label{eq:ScalarGNI}
{\cal L} \supset &~\overline{\nu} \nu~\left[ \overline{q}(C_{S}^{(q)}+D_{S}^{(q)}i\gamma^{5})q\right]  +  \frac{G_{F}}{\sqrt{2}}  ~    \overline{\nu}  i \gamma^{5}   \nu ~ \left[ \overline{q}(C_{P}^{(q)}    i \gamma^{5}  -  D_{P}^{(q)} )q\right]  \;,
 \end{align}
 where we have used $ D_{a}^{(q)}  =  \overline{D}_{a}^{(q)}  $ for $ a = S $ and $P$. The above Lagrangian can be expressed as
%
%%%%%
\begin{align} \label{eq:ChiralScalarGNI2}
 {\cal L}   \supset  &   ~\dfrac{1}{2} (C_S  + i D_P)^{}    \overline{\nu}  (1 - \gamma^{5})  \nu   ~   \overline{q} q  + \dfrac{1}{2} (C_S  - i D_P)^{}    \overline{\nu}  (1 + \gamma^{5})  \nu   ~   \overline{q} q \;.
\end{align}
We now define the scalar GNI, which are coefficient  of Eq. (\ref{eq:ChiralScalarGNI2}) as
 \begin{equation} \label{eq:ChiralScalarGNI3}
  \varepsilon_{S}  = \frac{1}{2}(C_S  + i D_P),   ~~~     \widetilde{\varepsilon}_{S}  = \frac{1}{2}(C_S  - i D_P)   \;.
 \end{equation}
%%%%%%%%%%%%%%%%
Similarly,  by collecting the coefficient of $ \gamma^{5} $ terms one can find pseudoscalar GNI.  In appendix \ref{app:GNI-Param}, we list all these GNI parameters in terms of   coefficients $C_{a}$ and $D_{a}$.
It is worth mentioning here the scalar GNI ($   \varepsilon_{S}  $) defined by Eq. (\ref{eq:ChiralScalarGNI3}) and that we have utilized later in our analysis  (see Eq.  (\ref{eq:GNIDefinition}) ) are equivalent up to a normalizing factor. 

As far as theoretical understanding is concerned, there exist many studies that explain vector-like GNI, for an example see Ref. \cite{Flores:2020lji}.
On the other hand,  leptoquark models as developed in \cite{Bischer:2019ttk} have addressed other GNI interactions like 
the scalar and tensor.
It can be seen from \cite{Bischer:2019ttk} that  the scalar and tensor interactions at dimension six occurs only for Dirac neutrinos, namely it is an interaction of left-handed with right-handed neutrinos.   For Majorana neutrinos, it is possible to have such an interaction at dimension seven,  which takes the form
\begin{equation}
{\cal L}_{O_7}\sim \bar{L} P_L L^{ } \bar{Q} u^{ } H \;.  
\end{equation}
Nevertheless, this operator contributes to $\Delta L=2$ processes and is very constrained by neutrinoless double beta decay~\cite{Bonnet:2012kh}.

%%%%%%%%%%%%%%%%%%%%%%%%%%%%%%%%%%%%%%%%%%%%%%%%%%%%%%%%%%%%%%%%%%%
%%%%%%%%%%%%%%%%%%%%%%%%%%%%%%%%%%%%%%%%%%%%%%%%%%%%%%%%%%%%%%%%%%%
\section{CE$\nu$NS Number of Events}\label{sec:coherent}
\subsection{In the Standard Model}\label{sec:SM}
%%%%%%%%%%%%%%%%%%%%%%%%%%%%%%%%%%%%%%%%%%%%%%%%%%%%%%%%%%%%%%%%%%%%

The SM weak-interaction differential cross-section for CE$\nu$NS as a function of the nuclear recoil energy $ E_r $, for a given nuclear mass $M_N$ is given by~\cite{Drukier:1983gj,Barranco:2005yy,Patton:2012jr}
\begin{equation}
\frac{d\sigma}{dE_r} = \frac{G_F^2}{2\pi}M_N Q_w^2 \left(2 - \frac{M_N E_r}{E_\nu^2}\right),
%\frac{d\sigma}{dE_r} = \frac{G_F^2}{4\pi}M_N Q_w^2 \left(1 - \frac{M_N E_r}{2 E_\nu^2}\right) F^2(Q^2),
\label{eq:crossSec}
\end{equation}
where $E_\nu$ represents the neutrino energy, and the weak nuclear charge can be read as
\begin{equation}
Q_w^2 = \left[ Z g^p_V F_p(Q^2) + N g^n_V F_n(Q^2)\right]^2 \;.
%Q_w = N - (1 - 4\sin^2\theta_W) Z .
\label{eq:weakCharge}
\end{equation}
Here, $Z$ and $N$ are the proton and neutron numbers, and  $F_{p,n}(Q^2)$ are the nuclear form factors  for a given momentum transfer $Q$ for protons and neutrons, respectively.  In this work, we consider the Klein-Nystrand form factor given by~\cite{Klein:1999qj}
\begin{equation}
F^{\rm KN}(Q^2) = 3 \frac{j_1 (Q R_A)}{Q R_A} \frac{1}{1+(Q a_k)^2}\;,
\end{equation}
where $j_1$ is the spherical Bessel function of the first kind, $a_k =0.7 $~fm, and $R_A = 1.23 \times A^{1/3}$ with $A$ being the nuclear mass number.

The  low-energy neutrino beam  generated at the Spallation Neutron Source (SNS) at the Oak Ridge National Laboratory consists of monochromatic $\nu_\mu$ coming from $\pi^+$ decays, $ \pi^+ \rightarrow \mu^{+} \nu_\mu$, along with delayed $\nu_e$ and $\bar{\nu}_\mu$ from the  subsequent $\mu^+$ decays, $ \mu^{+} \rightarrow e^+ \overline{\nu}_\mu \nu_e$. The flux components for $ \nu_\mu,  \nu_e$, and $ \bar{\nu}_\mu $ beams  are given by 
\begin{align}\label{eq:NuFlux}
	\phi_{\nu_\mu}(E_\nu) &= \eta \, \delta\left( E_\nu - \frac{m_\pi^2 - m_\mu^2}{2 m_\pi}\right), \nonumber \\
	\phi_{\nu_e}(E_\nu) &= \eta \frac{192 E_\nu^2}{m_\mu^3} \left(\frac{1}{2} - \frac{E_\nu}{m_\mu} \right), \\
	\phi_{\bar{\nu}_\mu}(E_\nu) &= \eta \frac{64 E_\nu^2}{m_\mu^3} \left(\frac{3}{4} - \frac{E_\nu}{m_\mu} \right), \nonumber
\end{align}
for neutrino energies $E_\nu \leq m_\mu / 2 \simeq 52.8$ MeV,  where  $\eta = r N_\mathrm{POT}/4\pi L^2$ is a normalization constant. Here,  $r $ is the fraction of neutrinos produced for each proton on target, $N_\mathrm{POT}$ represents the total number of protons on target ($2.1 \times 10^{23}$ POT per year), and $L$ is the distance to the detector~\cite{Akimov:2017ade}.

The differential recoil spectrum  is defined as
\begin{equation}
\frac{dR}{dE_r} = \sum_\alpha N_a \int_{E_\nu^\mathrm{min}}^{E_\nu^\mathrm{max}} \phi_\alpha (E_\nu) \frac{d\sigma}{dE_r} dE_\nu \;,
\label{eq:recoilSpec}
\end{equation}
where $N_a$ is the number of atoms in the detector, $M_\mathrm{det}$,  $\phi_\alpha (E_\nu)$ represents the neutrino flux for each flavor (see Eq.~(\ref{eq:NuFlux})), $E_\nu^\mathrm{min} = \sqrt{M_N E_r /2}$, and $E_\nu^\mathrm{max} = 52.8$~MeV.

In order to predict the number of events in both COHERENT detectors, quenching, energy smearing (or resolution), and detection efficiency must be taken into account.  First,  the nuclear recoil energy, $E_r$,  from the differential recoil spectrum in Eq.~\eqref{eq:recoilSpec} needs to be converted into the true electron-equivalent energy, $E_{ee}^{true}$, through the relation
	\begin{equation}
		E_{ee}^{true} = Q_f E_r \;,
	\end{equation} 
where $Q_f$ is the quenching factor. Then, the energy smearing should be applied in order to convert $E_{ee}^{true}$ to the reconstructed energy, $E_{ee}^{\,reco}$. Finally, the binned number of events can be computed by applying a detection efficiency function and integrating:
\begin{equation}
	N_i = \int_{E_r^i}^{E_r^{i+1}}A(E_{ee}^{\,reco})\frac{dR}{dE_{ee}^{\,reco}} dE_{ee}^{\,reco}\;.
	\label{eq:expectedEvents}
\end{equation}

For the case of the CsI detector, quenching can be applied with the following function
\begin{equation}
	E_{ee}^{true} = a E_r + b E_r^2 + c E_r^3 + dE_r^4 \;, 
\end{equation}
where the coefficients $a,b,c$ and $d$ are provided in~\cite{Akimov:2021dab}. For quenched recoils in the CsI detector, the light yield is given by $L_Y = 13.35$ PE/keV$_{ee}$. The energy smearing is applied through the following Gamma distribution
\begin{equation}
	P(E_{ee}^{true},E_{ee}^{reco}) = \frac{(a(1+b))^{1+b}}{\Gamma(1+b)} (E_{ee}^{reco})^b e^{-a(1+b)E_{ee}^{reco}},
\end{equation}
where $a=1/(L_Y E_{ee}^{true})$ and $b=0.7157 L_Y E_{ee}^{true}$. The energy and time efficiency functions are also specified in~\cite{Akimov:2021dab}. Following this procedure we found 336 events in the SM framework.

For the LAr detector, the quenching factor is parameterized as $Q_f = (0.246) + 7.8\times 10^{-4}E_r ~\mbox{keV}_{nr}$~\cite{Akimov:2020czh}. Energy smearing must be applied with a Gaussian energy resolution function with $\sigma / E_{ee}^{\,reco} = 0.58 / \sqrt{E_{ee}^{\,reco} (\mbox{keV}_{ee})}$. The detection efficiency is also provided in~\cite{Akimov:2020czh}.
Following this procedure we can study the effects of new physics scenarios like GNI, by comparing with the latest COHERENT measurements.

%%%%%%%%%%%%%%%%%%%%%%%%%%%%%%%%%%%%%%%%%%%%%%%%%%%%%%%%%%%%%%

%%%%%%%%%%%%%%%%%%%%%%%%%%%%%%%%%%%%%%%%%%%%%%%%%%%
%%%%%%%%%%%%%%%%%%%%%%%%%%%%%%%%%%%%%%%%%%%%%%%%%%%%%%%%%%%%%%%%%%%
\subsection{In presence of  GNI}\label{sec:GNI-Events}
%%%%%%%%%%%%%%%%%%%%%%%%%%%%%%%%%%%%%%%%%%%%%%%%%%%%%%%%%%%%%%%%%%%%
In order to examine the role of GNI  at CE$\nu$NS process,  one needs to find how the differential cross-section changes in presence of GNI.  
 Following Ref.~\cite{Lindner:2016wff, AristizabalSierra:2018eqm}, the differential cross-section in presence of GNI at leading order is
\begin{align}\label{eq:CS-GNI}
	\left( \frac{d\sigma}{dE_r}\right) ^{f} &= \frac{G_F^2}{4\pi}M_N N^2 F^2(Q^2)  \times  \nonumber \\
	 & \Bigg[\xi^{f\,2}_S \frac{E_r}{E^\mathrm{max}_r}
 + (\xi^f_V + A_{\rm SM})^2 \left(1-\frac{E_r}{E^\mathrm{max}_r}-\frac{E_r}{E_\nu}\right)  \pm 2\xi^f_V \xi^f_A\frac{E_r}{E_\nu} \nonumber \\
 &+ \xi_A^{f\,2} \left(1+\frac{E_r}{E^\mathrm{max}_r}-\frac{E_r}{E_\nu}\right) + \xi^{f\,2}_T \left(1-\frac{E_r}{2 E^\mathrm{max}_r}-\frac{E_r}{E_\nu}\right) \mp  R \frac{E_r}{E_\nu} + \mathcal{O} \left(\frac{E^2_r}{E^2_\nu}\right)\Bigg] \;,
\end{align}
where  the SM contribution is $A_{\rm SM} = 1 - (1- 4 \sin^2 \theta_w)Z/N$, $ E_r  $   is the recoil energy of the nucleus  with $ E^\mathrm{max}_r = 2 E^{2}_\nu/ M_N$,  and 
\begin{align}\label{eq:GNIDefinition}
&\xi_{S}^{f\,2}=\frac{1}{N^{2}}(C_{S}^{2}+D_{P}^{2}),~~\thinspace  \xi_{V}^f=\frac{1}{N}(C_{V}-D_{A}), \nonumber \\
 &\xi_{A}^f=\frac{1}{N}(C_{A}-D_{V}),~~ \thinspace  \xi_{T}^{f\,2}=\frac{8}{N^{2}} (C_{T}^{2} + D_T^2), \\
  &R = \dfrac{2}{N^{2}} (C_S C_T - C_PC_T + D_S D_T - D_P D_T).   \nonumber
\end{align}
Here, $\xi_X$'s (for $ X = S,  V, A, T $) represent scalar ($ S $), vector  ($ V $), axial-vector  ($ A $), and tensor  ($ T $) interactions for a specific lepton flavor $f$, 
respectively, and $R$ is an interference term.  
To analyze the importance of these GNI terms individually, we do not consider any interference terms in our study.
Also, one can recover the SM limit by setting $ \xi_{S},  \xi_V, \xi_A, \xi_T  = 0 $.   
We clarify here that the axial-vector current and interference term $R$  of Eq.~\eqref{eq:CS-GNI}  are the two term that come with opposite signs in the $ \nu^{}(\overline{\nu}) - N $ cross-section. Since, here we are not examining the impact of both these terms, in our numerical analysis we can safely assume equal cross-section for both neutrinos and anti-neutrinos.

%%%%%%%%%%%%%%%%%%%%%%%%%%%%%%%%%%%%%%%%%%%%%%%%%%%%%
\begin{figure}[t]
	\includegraphics[width=0.49\textwidth]{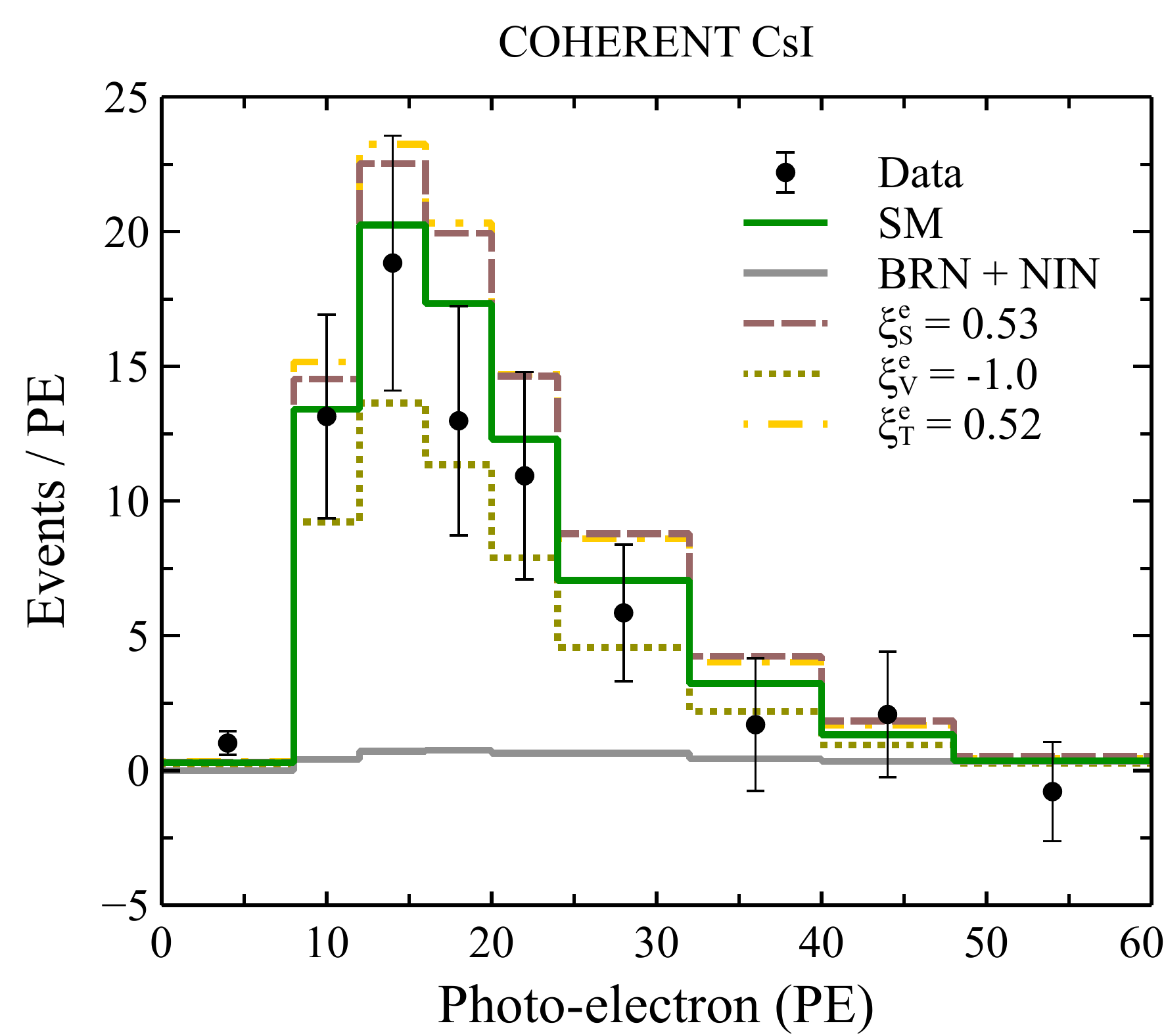}%
	\hspace{0.2cm}
	\includegraphics[width=0.49\textwidth]{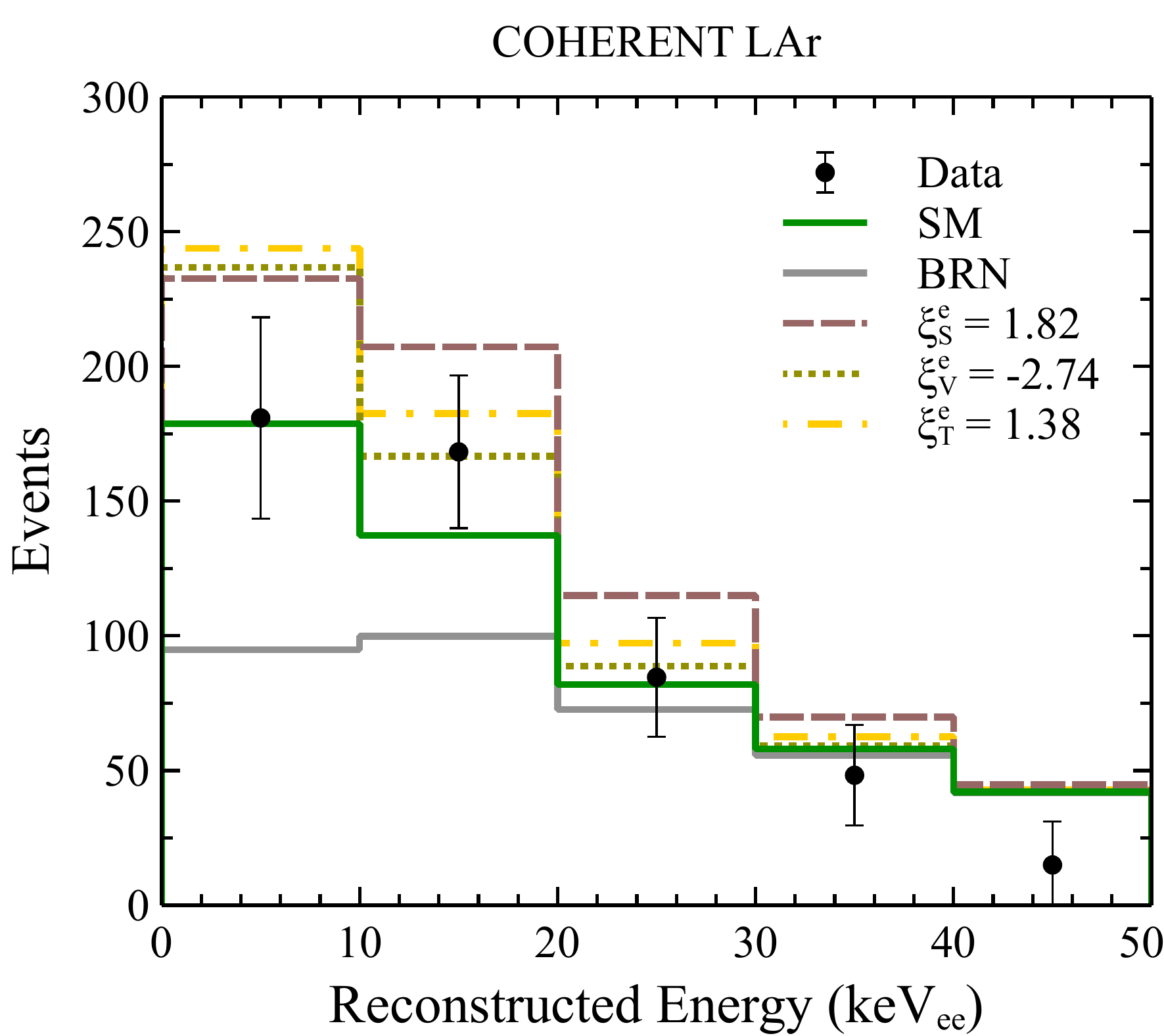}
	\caption{\footnotesize Expected number of events for the CsI (left panel), and LAr (right panel) detectors of the COHERENT collaboration.  The SM predictions are shown using the solid-green histogram, whereas the dashed,  dotted, and dashed-dotted histograms  replicate calculated events in presence of  three-benchmark values of different GNI parameters. We also show with black dots the most recent measurements along with their uncertainty. In the left panel, all histograms represent the number of events divided by the bin size, whereas in the right panel they represent the total number of events plus background. }
	\label{fig:CsI-GNI_Events_Ele}
\end{figure}
%%%%%%%%%%%%%%%%%%%%%%%%%%%%%%%%%%%%%%%%%%%%%%%%%%%%%

We now present expected number of events in Fig. \ref{fig:CsI-GNI_Events_Ele}.
In the left panel, we show the latest COHERENT spectral data as a function of photo-electron (PE) for the CsI detector using the black cross marks~\cite{Akimov:2021dab}.
Similarly, we show the spectral data for the LAr detector as a function of the reconstructed Energy $({\rm keV_{ee} })$ in the right panel~\cite{Akimov:2020czh}.
 The SM predictions are obtained using the event number expression of Eq. \eqref{eq:expectedEvents} along with  the differential cross section as defined by  Eq. \eqref{eq:crossSec}.  
Later,  in order to examine the impact of GNI on CE$\nu$NS, we calculate expected number of events for three benchmark values of  GNI (see figures for the detailed benchmark values), where we use the modified SM differential cross-section as given by Eq.~\eqref{eq:CS-GNI}.
Note that these benchmark  values are not arbitrary rather taken from $ 90\%$ C. L. range of our $ \chi^{2} $ analysis (see Table \ref{tab:GNI-BenchMark}),  which we discuss in details in subsequent section.

\section{Chi-square analysis}\label{sec:Nu-Analysis}
%%%%%%%%%%%%%%%%%%%%%%%%%%%%%%%%%%%%%%%%%%%%%%%%%%%%%%%%%%%%%%
In order to constrain the GNI parameters $\xi^f_X$ using the available experimental data from COHERENT, we perform a $\chi^2$ analysis considering different scenarios. First, a one-parameter analysis has been done for each detector specification, where we set all $\xi^f_X$ equal to zero, except for one non-zero GNI at a time. In this way, we clearly see the impact of the current measurements on each Lorentz structure ($S,V,T$) and for each flavor ($e,\mu$).
Then, we perform a two-parameter fit, where we adopt two non-zero GNI parameters at a time, with the same Lorentz structure but different flavor, i.e.\ $\xi_X^e-\xi_X^\mu$. Finally, a  combined analysis has been done considering both the detectors. 

For the  COHERENT-CsI data, we define the following $ \chi^{2} $ function
%%%%%%%
\begin{equation}
\chi^2_\mathrm{CsI} = \sum_{i=2}^{8} \left[\frac{N_\mathrm{meas}^i - (1+\alpha)N_\mathrm{th}^i  (\xi_a^f) - (1+\beta)B_\mathrm{on}^i}{\sigma_\mathrm{stat}^i} \right]^2 + \left(\frac{\alpha}{\sigma_\alpha}\right)^2 + \left(\frac{\beta}{\sigma_\beta}\right)^2 \;,
\label{eq:chiSqFunCsI}
\end{equation}
where $N_\mathrm{meas}^i$ is the measured number of events per energy bin,   $ N_\mathrm{th}^i  (\xi_a^f) $  is the expected number of events in presence of GNI  per energy bin, and $B_\mathrm{on}^i$ is the beam-on background.
The statistical uncertainty  is defined as $\sigma_\mathrm{stat}^i = \sqrt{N_\mathrm{meas}^i + B_\mathrm{on}^i + 2B_\mathrm{ss}^i}$, where $B_\mathrm{ss}^i$ represents the steady-state backgrounds.
The total systematic uncertainty in the signal is represented as $\sigma_\alpha$, which includes uncertainties from signal acceptance (4.1\%), neutrino flux (10\%), nuclear form factor (3.4\%), and the most recent COHERENT-2020 measurement of the quenching factor (3.8\%)~\cite{COHERENT:2021pcd}.  Adding these in quadrature, we have $\sigma_\alpha = 0.1195$.
On the other hand, the systematic uncertainty for the background has the value $\sigma_\beta=0.25$. The function in Eq.~\eqref{eq:chiSqFunCsI} must be minimized over the nuisance parameters $\alpha$ and $\beta$.

For the analysis of the CENNS-10 detector data, we adopt \textit{analysis A} of Ref.~\cite{Akimov:2020pdx} and consider the  following $\chi^2$ function
\begin{eqnarray}
\chi^2_\mathrm{LAr}
&=&
\sum_{i=1}^{3}
\left(
\frac{
N^{i}_\mathrm{meas} - (1+\alpha) N^i_\mathrm{th}(\xi_a^f)-(1+\beta) B^i_\mathrm{PBRN}-(1+\gamma)B^i_\mathrm{LBRN}}{\sigma^i_\mathrm{stat}}
\right)^2\\ \nonumber
&+&
\left( \frac{\alpha}{\sigma_\alpha} \right)^2
+
\left( \frac{\beta}{\sigma_\beta} \right)^2
+
\left( \frac{\gamma}{\sigma_\gamma} \right)^2
,
\label{eq:chiSqFunLAr}
\end{eqnarray}
where the subscript PBRN (LBRN) stands for Prompt (Late) Beam-Related Neutron background, and $(\sigma^i_\mathrm{stat})^2 = \left( \sigma^i_\mathrm{meas} \right)^2 + \left[ \sigma_{\mathrm{BRNES}} \left( B^i_{\mathrm{PBRN}} + B^i_{\mathrm{LBRN}}\right)\right]^2$, with $\sigma^i_\mathrm{meas}$ and $\sigma_\mathrm{BRNES}$ being the statistical uncertainty from CE$\nu$NS events and from Beam Related Neutron Energy Shape, respectively. The systematic uncertainties from BRN background take the value $\sigma_\beta = 0.32$ and $\sigma_\gamma=1.0$~\cite{Akimov:2020czh}, while for the remaining ones we adopted the estimated values $\sigma_\alpha=0.134$ and $\sigma_\mathrm{BRNES} = 0.017$~\cite{Cadeddu:2020lky}.
The function in Eq.\eqref{eq:chiSqFunLAr} is similar to the one adopted in Ref.~\cite{Cadeddu:2020lky}, but with the main difference that we consider only the first three energy bins (see Appendix~\ref{app:LAr_analyses} for a comprehensive discussion on different approaches for the $\chi^2$ definition).

\subsection{One and two-parameter analyses}\label{sec:OneParam}
%%%%%%%%%%%%%%%%%%%%%%%%%%%%%%%%%%%%%%%%%%%%%%%%%%%%%
%%%%%%%%%%%%%%%%%%%%%%%%%%%%%%%%%%%%%%%%%%%%%%%%%%%%%
\begin{figure} [t]
	\includegraphics[width=\textwidth]{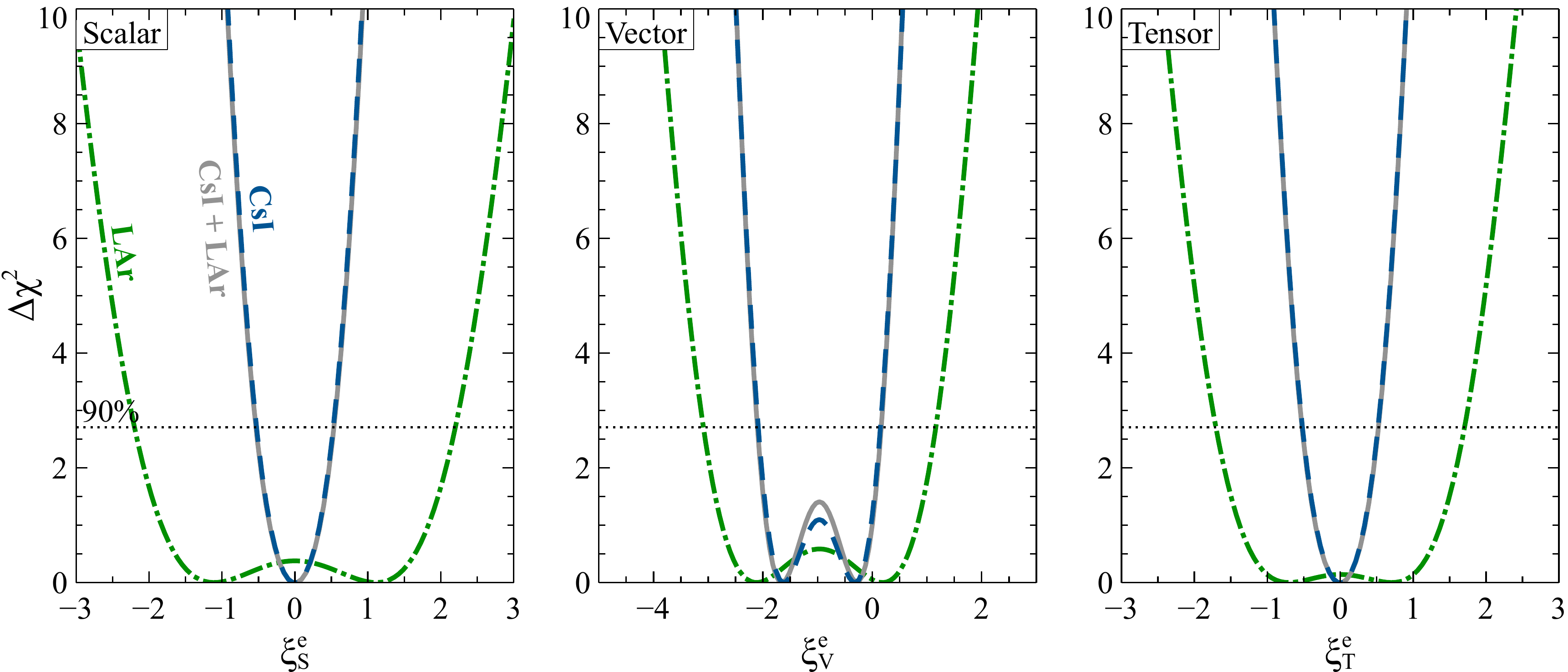} \\
	\vspace{0.5cm}
	\includegraphics[width=\textwidth]{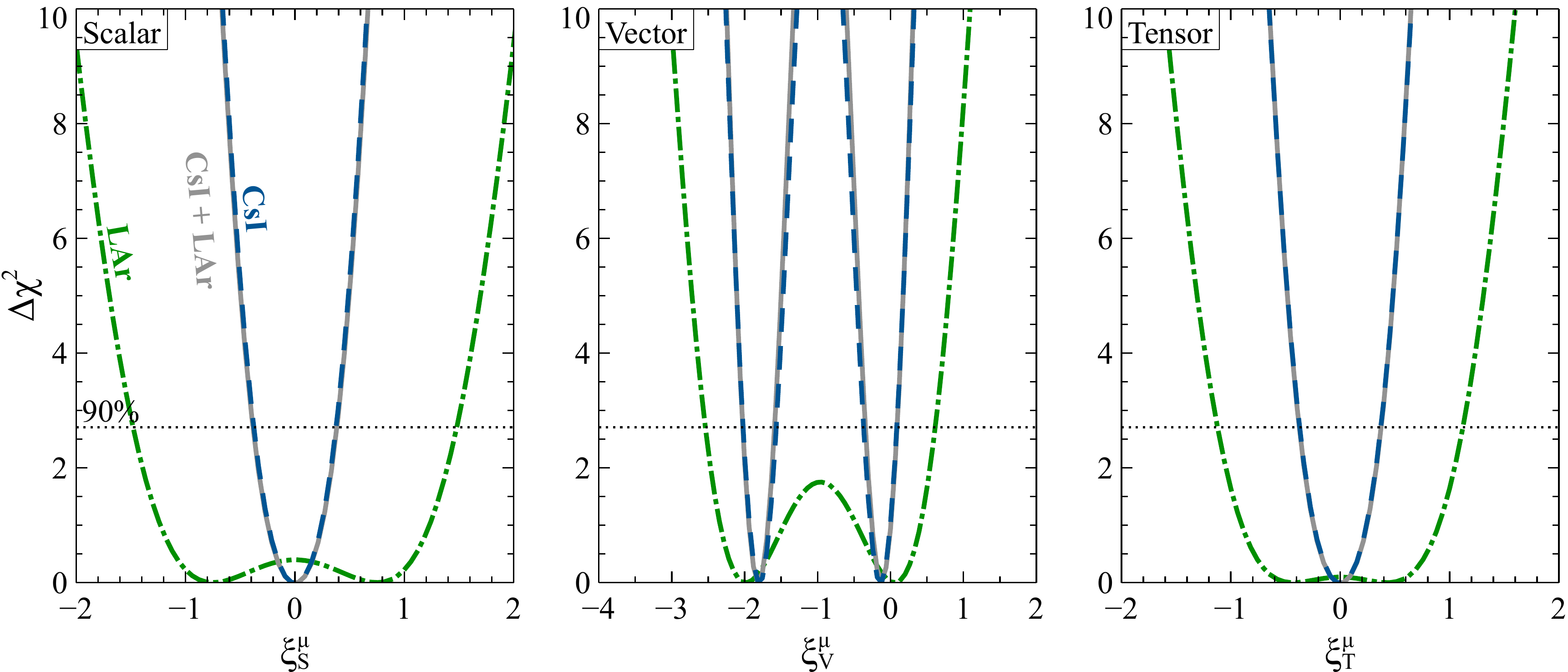}
	\caption{\footnotesize Constraints on the scalar, vector, and tensor flavored GNI parameters from the COHERENT CsI and LAr measurements (dashed blue and dotted-dashed green curves, respectively), and their combination (solid gray curve). The upper (lower) panels show the limits on electron (muon) flavor GNI parameters.}
	\label{fig:Chisq_one}
\end{figure}
%%%%%%%%%%%%%%%%%%%%%%%%%%%%%%%%%%%%%%%%%%%%%%%%%%%%%

Having described the statistical approach for both COHERENT data sets, we start with a one-parameter analysis: we take only one $\xi_X^f$ different from zero at the time. Since we are studying only scalar, vector, and tensor interactions, and given that the SNS produces neutrinos with electron and muon flavor~\footnote{Since the cross-section for GNI is the same for neutrinos and antineutrinos considering only one interaction at the time (neglecting axial terms), we can take the same coupling for $\nu_\mu$ and $\bar{\nu}_\mu$.}, our analysis considers only six different parameters.
  The results are shown in Fig.~\ref{fig:Chisq_one}, where the upper (lower) panels correspond to the different GNI electron (muon) couplings. 

From Fig.~\ref{fig:Chisq_one} we can see that the electron couplings are less constrained than muon couplings given the lower $\nu_e$ flux at the SNS. We can also notice that the limits from CsI measurement are better than the LAr data at constraining every GNI parameter. The main reasons for this are the larger background on the LAr detector and  the recent improvement in the measurement of the quenching factor  and the increased statistics of the CsI detector. Since the measured number of events in the CsI detector is less than the SM expectation, scalar and tensor interactions are more constrained and their best-fit value is at zero, given that these interactions can only add events to the overall signal. On the contrary, the LAr detector measured more events than the expected from the SM, hence scalar and tensor interactions are less constrained and their best-fit values are different from zero \footnote{There are two minima for these interactions given the quadratic dependence of each coupling in Eq.~\eqref{eq:CS-GNI}. }.

Now looking at the vector GNI case, we find two minima  for both detectors (more evident in $\xi_V^\mu$). This degeneracy can be explained by looking at the coefficient of the second term of Eq.~\eqref{eq:CS-GNI}. Since the spectral shape of both CsI and LAr measurements resembles the vector interaction, a value of $\xi_V\sim 0$ fits the data (only SM contribution); in the same way, a value of $\xi_V\sim -2$ resembles the measured signal due to a partial destructive interference with the SM value.

For combined analysis, we notice from both the rows of Fig.~\ref{fig:Chisq_one} (see solid gray curves) that the results are nearly identical to the CsI analysis alone. This is because the allowed regions for the CsI case are more stringent than the ones from the LAr case, hence the latter have an almost negligible effect in the combined region. 

%%%%%%%%%%%%%%%%%%%%%%%%%%%%%%%%%%%%%%%%%%%%%%%%%%%%%
\begin{table}
	\begin{tabular}{|c|c|c|c|c|}
		\hline
		& \multicolumn{2}{c|}{This work} & \multicolumn{2}{c|}{Previous works}  \\
		\hline
		& CsI & LAr & CsI~\cite{Han:2020pff} & CsI~\cite{AristizabalSierra:2018eqm} \\
		\hline
		$\xi_S^e$	  & [-0.53, 0.53] & [-2.19, 2.19]   & [-1.22, 1.22]  & \multirow{2}{*}{[-0.62, 0.62]} \\
		$\xi_S^\mu$	& [-0.37, 0.37] & [-1.48, 1.48]   & [-0.77, 0.77] &  \\
		\hline
		$\xi_V^e$     & [-2.06, 0.15]   & [-3.08, 1.16]   &  --&   \multirow{2}{*}{[-2.102, -1.554] \& [-0.324, 0.224]} \\
		$\xi_V^\mu$ & [-2.02, -1.56] \& [-0.37, 0.09]  & [-2.53, 0.61]    & -- &  \\
		\hline
		$\xi_T^e$     & [-0.52, 0.52] & [-1.71, 1.71]    & [-1.26, 1.26] &  \multirow{2}{*}{[-0.591, 0.591]}\\
		$\xi_T^\mu$ & [-0.37, 0.37] & [-1.11, 1.11]   & [-0.85, 0.85] &  \\
		\hline
	\end{tabular}
	\caption{\footnotesize Allowed regions of the scalar, vector, and tensor GNI at  90\% C. L. from the COHERENT-CsI and LAr detector analyses. The first two columns correspond to the results of this work, while the other columns to the limits obtained in~\cite{Han:2020pff} and~\cite{AristizabalSierra:2018eqm}. The limits from the last column where obtained assuming $\xi_X^e = \xi_X^\mu$. 
		%The values for LAr are obtained with the binned ROI $\Delta\chi^2$.}
	}
	\label{tab:GNI-BenchMark}
\end{table}
%%%%%%%%%%%%%%%%%%%%%%%%%%%%%%%%%%%%%%%%%%%%%%%%%%%%%

 We present the 90\% C. L. regions for all six $\xi_X^f$ parameters for both CsI and LAr measurements 
in Table~\ref{tab:GNI-BenchMark}. For the sake of completeness, we also show the limits obtained in other works using the old CsI data.  In Ref.~\cite{Han:2020pff}, the authors studied flavored scalar and tensor interactions, whereas in Ref.~\cite{AristizabalSierra:2018eqm}, the authors studied the same Lorentz structure GNI parameters of this work, but without assuming a flavor dependence. The $\xi_X^e = \xi_X^\mu$ assumption strengthens the limits, but the use of the old quenching factor weakens them. 

\begin{figure} [t]
	\includegraphics[width=\textwidth]{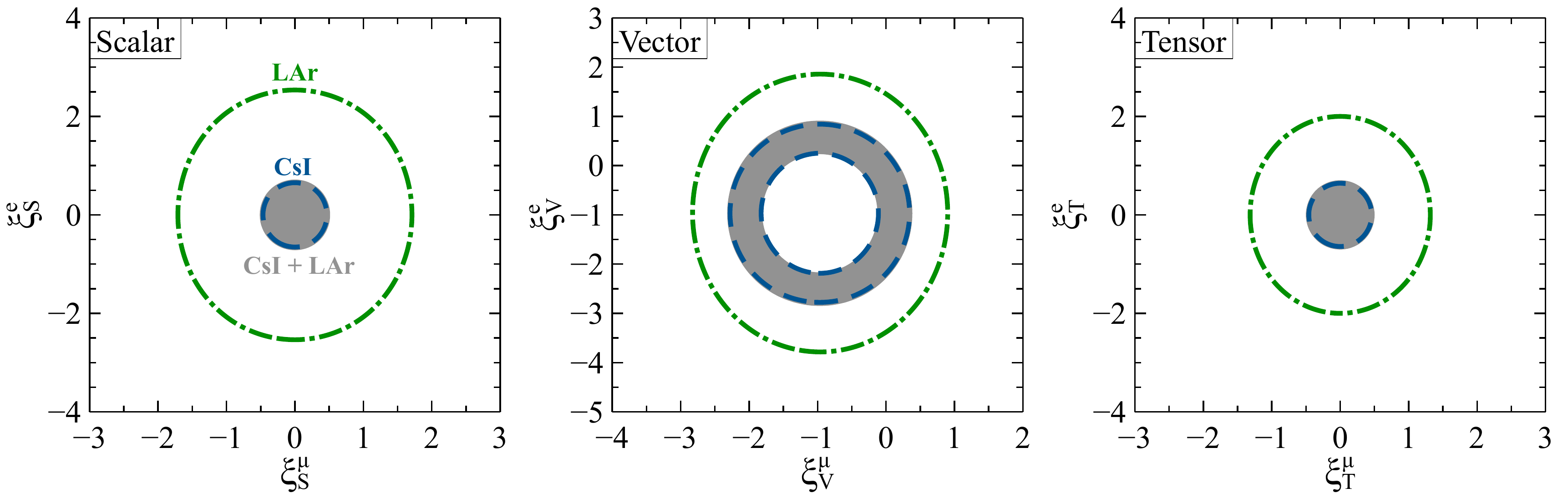}
	\caption{\footnotesize Allowed regions at 90\% C. L. on the $\xi_X^e - \xi_X^\mu$ flavored GNI parameter plane. The regions limited by the dashed-blue and dotted-dashed green curves arises from  the COHERENT-CsI and LAr measurements, respectively. Their combined limit is represented by the gray region.}
	\label{fig:Chisq_two}
\end{figure}

Having studied the case with only one non-zero GNI parameter at a time,  we now proceed to discuss the case with two non-zero parameters. The interplay between two different types of Lorentz structures has been studied, for example, in~\cite{AristizabalSierra:2018eqm} and~\cite{Han:2020pff}. On the contrary, in this work, we focus on the correlation between parameters with the same structure ($S,V,T$) but with different lepton flavors ($e,\mu$).
Our results at $90\%$ C. L. are presented in Fig.~\ref{fig:Chisq_two}, using the same color-code as in the one-parameter case. Again, we can see how tensor interactions are the more constrained by the COHERENT measurements.
A continuous degenerate region has been observed for the COHERENT-CsI  and the combined study, as is apparent from the one-parameter analysis of the middle panels of Fig.~\ref{fig:Chisq_one}.
In order to break these continous degenerate solutions, in the next section we combine the COHERENT data  with the projected measurements of reactor CE$\nu$NS experiments. 

%%%%%%%%%%%%%%%%%%%%%%%%%%%%%%%%%%%%%%%%%%%%%%%%%%%%%
%%%%%%%%%%%%%%%%%%%%%%%%%%%%%%%%%%%%%%%%%%%%%%%%%%%%%

%%%%%%%%%%%%%%%%%%%%%%%%%%%%%%%%%%%%%%%%%%%%%%%%%%%%%
\section{Projected sensitivities from the SBC-CE$\nu$NS detector}\label{sec:SBC}
%%%%%%%%%%%%%%%%%%%%%%%%%%%%%%%%%%%%%%%%%%%%%%%%%%%%%
As already mentioned, there are currently several ongoing (and/or proposed) reactor-based CE$\nu$NS experiments aiming to measure this process for the first time. One of these proposals is the SBC-CE$\nu$NS detector, currently under construction by the SBC collaboration~\cite{Flores:2021ihw}. This detector is a liquid-noble scintillating bubble chamber capable of achieving a 100 eV energy threshold and a great background reduction given its blindness to electron recoils. Currently, the SBC collaboration is building a 10-kg LAr chamber which is planned to be set at 3 m from the 1-MW$_{th}$ TRIGA Mark III research reactor at the National Institute for Nuclear Research (ININ). A detailed discussion of this experiment and its reach for the SM and beyond the SM scenarios can be seen in~\cite{Flores:2021ihw}.

In order to study the projected sensitivity of this detector to the interactions described in this work, we will consider the following $\chi^2$ function
\begin{equation}
	\chi^2 =\left[\frac{N_\mathrm{meas} - (1+\alpha)N_\mathrm{th}(\gamma)- (1+\beta)B_\mathrm{reac}}{\sigma_\mathrm{stat}}\right]^2 + \left(\frac{\alpha}{\sigma_\alpha}\right)^2 + \left(\frac{\beta}{\sigma_\beta}\right)^2 + \left(\frac{\gamma}{\sigma_\gamma}\right)^2,
	\label{eq:chisq_SBC}
\end{equation}
where $N_\mathrm{th}$ the number of events including GNI, $N_\mathrm{meas}$ is the SM prediction, and $\sigma_\mathrm{stat}=\sqrt{N_\mathrm{meas} + 4 B_\mathrm{cosm}}$ its statistical uncertainty. Background from the reactor and from muon-induced and cosmogenic neutrons are represented by $B_\mathrm{reac}$ and $B_\mathrm{cosm}$, respectively. The nuisance parameters $\alpha, \beta$ and $\gamma$ account for the uncertainties on the signal, background, and threshold, respectively. These uncertainties take the values $\sigma_\alpha=0.024$, $\sigma_\beta=0.1$, and $\sigma_\gamma=0.05$. The nuclear recoil threshold is taken as (1+$\gamma$)$\cdot$100~eV.

%%%%%%%%%%%%%%%%%%%%%%%%%%%%%%%%%%%%%%%%%%%%%%%%%%%%%
\begin{figure} [t]
	\includegraphics[width=\textwidth]{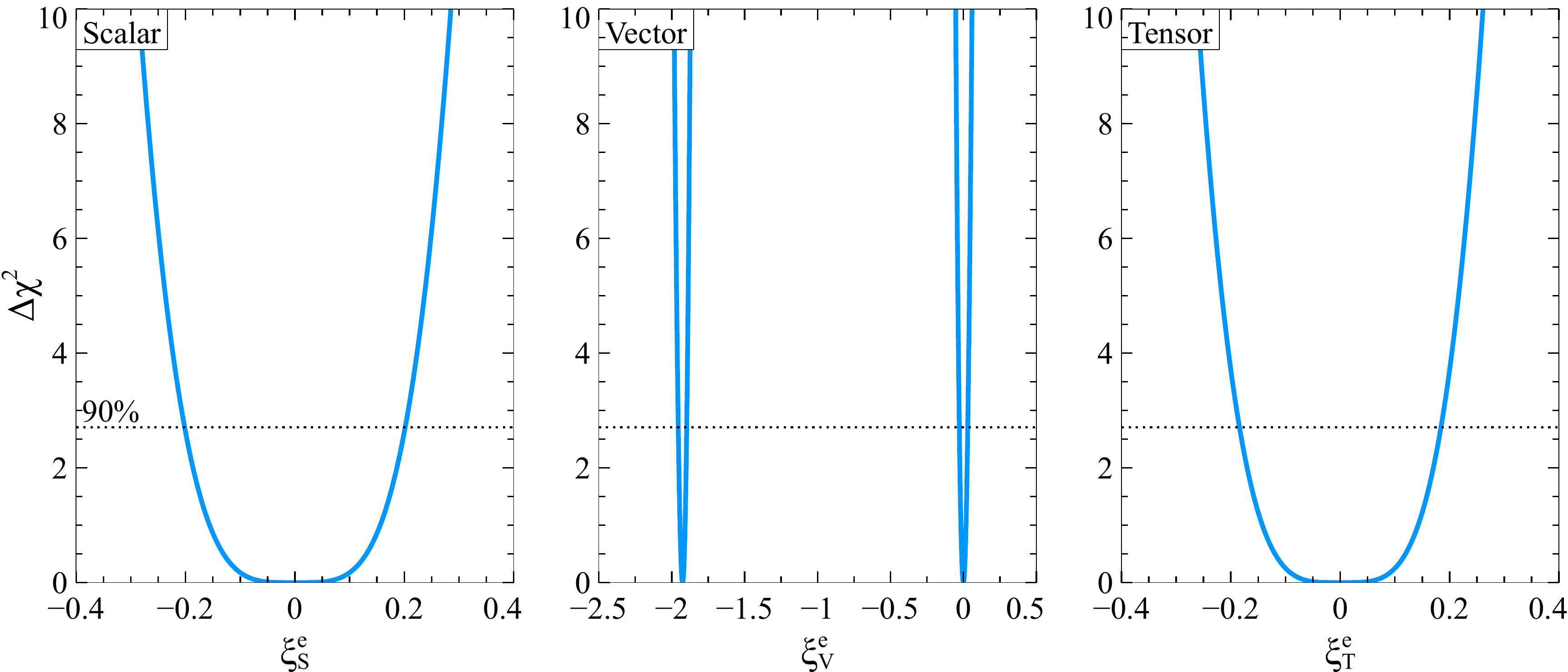}
	\caption{\footnotesize Projected constraints on the scalar, vector, and tensor flavored GNI parameters from the upcoming  SBC-CE$\nu$NS detector.}
	\label{fig:Chisq_SBC}
\end{figure}
%%%%%%%%%%%%%%%%%%%%%%%%%%%%%%%%%%%%%%%%%%%%%%%%%%%%%

By following the same approach as in the previous section, we derive limits on the GNI parameters, one at a time. Our results are shown in Fig.~\ref{fig:Chisq_SBC}. Let us remark that at a nuclear reactor, only electron antineutrinos are produced, hence we present limits only for the $\xi^e_{X}$ parameters.
Since we are assuming that the SBC-CE$\nu$NS experiment will measure the SM predictions, the allowed values for a new vector interactions are more constrained than for the scalar and tensor interactions.
It can be observed that these projected constraints are stronger than the current COHERENT limits due to the intense $\bar{\nu}_e$ flux, and the expected low threshold and highly-reduced background of the SBC-CE$\nu$NS experiment.

%%%%%%%%%%%%%%%%%%%%%%%%%%%%%%%%%%%%%%%%%%%%%%%%%%%%%
\subsection{Combined COHERENT and SBC-CE$\nu$NS analysis}
\begin{figure}[t]
	\includegraphics[width=\linewidth ]{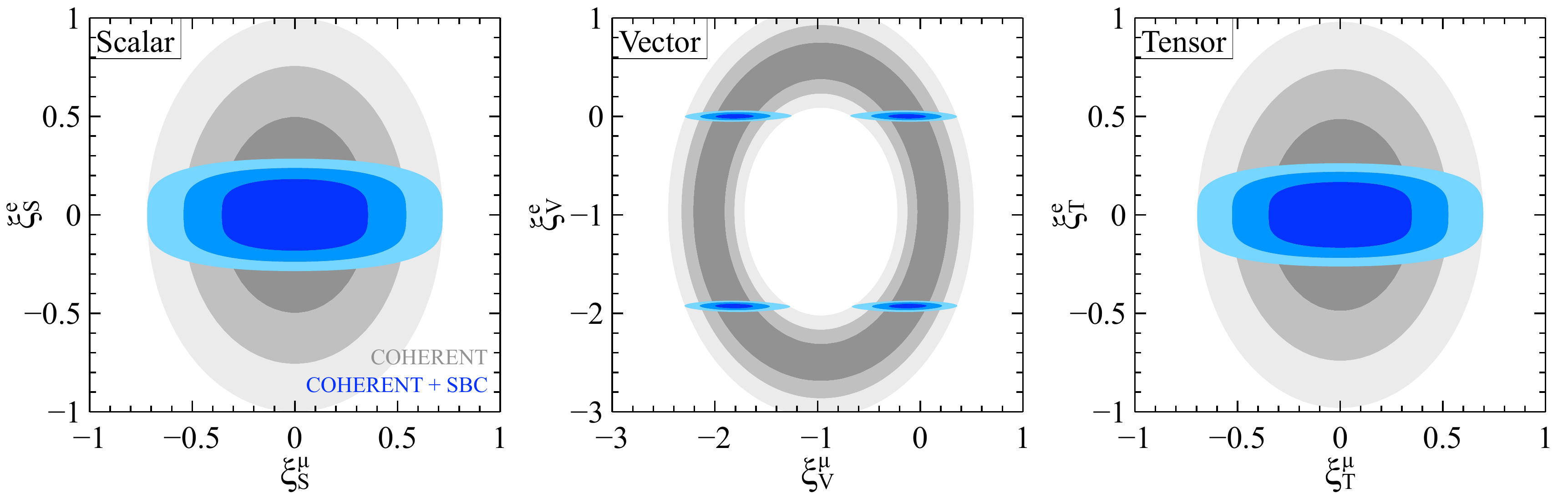}
	\caption{\footnotesize Allowed regions for the scalar, vector and tensor GNI parameters at $1\sigma$, $2\sigma$, and $3\sigma$ CL, in the $\xi_{X}^\mu - \xi_{X}^e$ plane. The gray-scale regions correspond to the limits obtained from the COHERENT measurements, while the blue-scale regions arise from the combination of COHERENT and the expected sensitivities of the SBC-CE$\nu$NS detector. }
	\label{fig:combinedSBC}
\end{figure}

%%%%%%%%%%%%%%%%%%%%%%%%%%%%%%%%%%%%%%%%%%%%%%%%%%%%%
As a final step in our GNI analysis in CE$\nu$NS experiments, we combined the current COHERENT data with the upcoming sensitivities of the SBC-CE$\nu$NS detector. Since the latter will be sensitive only to the $\xi^e_{X}$ parameters, we have decided to perform an analysis by taking two parameters different from zero at a time. The allowed region of this combined analysis, at $1, 2$, and $3^{}\sigma$ C. L. are displayed in Fig.~\ref{fig:combinedSBC}. For a comparison, we also show the allowed parameter space for the combined COHERENT data.
It has been observed that the allowed regions are tightly constrained for the electron flavor due to the inclusion of simulated SBC-CE$\nu$NS data. The most remarkable result is expected in the vector interactions, where we notice four-fold discrete degeneracies (see blue contours in the middle panel).

%%%%%%%%%%%%
\section{Conclusions}\label{sec:Conclusion}
%%%%%%%%%%%%
In this work, we have performed an analysis to examine generalized neutrino interactions (GNI) in light of the recent COHERENT-CsI  and LAr data. This work is concentrated on electron and muon flavored GNI. In doing so, a chi-square analysis has been performed to constrain these GNI parameters. Later, we have also added the simulated data arising from the planned SBC-CE$\nu$NS experiment.

We summarize our noteworthy results in Figs.~\ref{fig:CsI-GNI_Events_Ele}- \ref{fig:combinedSBC}, and Table \ref{tab:GNI-BenchMark}.   The expected number of events for both the COHERENT data sets are described in Fig.  \ref{fig:CsI-GNI_Events_Ele}.  For comparison, we have shown the measured SM events as well as the expected number of events for the  scalar, vector, and tensor interactions, respectively. 
From one-parameter $ \chi^{2} $ analysis, we have noticed from Fig.  \ref{fig:Chisq_one} that  $ \xi^\mu_{X} $'s are stronger constrained compared to $ \xi^e_{X} $'s.  This is due to the lower $\nu_e$ flux at the SNS than that of  the $\nu_\mu$ flux.
It has also been observed that the scalar and tensor GNI prefer their best-fit values at zero for the CsI detector, whereas a degeneracy at  the $ \chi^{2} $ minimum has been identified for LAr data.  We have noticed further that  the combined analysis of CsI and LAr data lift these degeneracies and best-fit values  have been observed at zero.
For the vector-GNI, we have found degeneracies at $ \xi_V =0, -2$ and which is due to the presence of interference term  as can be seen from Eq.~\eqref{eq:CS-GNI}.
We have listed one-parameter $ \chi^{2} $ at 1$ \sigma $ confidence level in Table \ref{tab:GNI-BenchMark}.

From the combined analysis of COHERENT data, it has been observed that (see Fig.~\ref{fig:Chisq_one}, and \ref{fig:Chisq_two}) the allowed parameter space of GNI parameters remain almost the same as in the case of the CsI detector alone.
Finally, we have found that the addition of   SBC-CE$\nu$NS  data significantly improved our results due to its intense flux, low threshold, and highly-suppressed background. 
It can be noticed from  Fig.  \ref{fig:combinedSBC} that electron-flavored GNI get severely constrained due to anti-neutrino flux arising from reactor data. Noticeably, a four-fold discrete degenerate solutions have been identified for the vector-like GNI.

%%%%%%%%%%%%%%%%%%%%%%%%%%%%%%%%%%%%%%%%%%%
\acknowledgments
%%%%%%%%%%%%%%%%%%%%%%%%%%%%%%%%%%%%%%%%%%%
We thank Daniel Pershey for all the clarifications regarding the computation of the expected number of events for the new COHERENT-CsI data. We also thank Omar Miranda and Gonzalo S\'anchez for the helpful discussion. This work is supported by 
the  German-Mexican  research  collaboration grant SP 778/4-1 (DFG) and 278017 (CONACYT),  
the grants CONACYT CB-2017-2018/A1-S-13051 (M\'exico), the DGAPA UNAM grant PAPIIT IN107621 and SNI (M\'exico). The work of N.N. is supported by the postdoctoral fellowship program DGAPA-UNAM and  partly supported by the Istituto Nazionale di Fisica Nucleare (INFN) through the “Theoretical Astroparticle Physics” (TAsP) project.
%%%%%%%%%%%%%%%

%%%%%%%%%%%%%%%%%%%%%%%%%%%%%%%%%%%%%%%%%%%
\appendix
\section{ GNI Parametrization}\label{app:GNI-Param}
%%%%%%%%%%%%%%%%%%%%%%%%%%%%%%%%%%%%%%%%%%%

In literature,  there exists two different parametrizations for GNI
interactions~\cite{Bischer:2019ttk}.   In this appendix, we establish relation between these two parametrizations.   The  effective Lagrangian in terms of C and D parametrization can be written as
\begin{equation}
{\cal L}\supset   \frac{G_{F}}{\sqrt{2}}   \sum_{a=S,P,V,A,T}\overline{\nu}\,\Gamma^{a}\nu\left[\overline{q}\Gamma^{a}(C_{a}^{(q)}+\overline{D}_{a}^{(q)}i\gamma^{5})q\right]\;,   \label{eq:Lagrangian-GNI-1}
\end{equation}
where the five possible  Dirac $ \Gamma^{a} $ matrices  are given by
\begin{equation}
	\Gamma^{a}\in\{1,i\gamma^{5},\gamma^{\mu},\gamma^{\mu}\gamma^{5},\sigma^{\mu\nu}\}\,.
\end{equation}
The relations between the coefficients $C_{a}^{(q)}$ and 
\begin{equation*}
	D_{a}^{(q)}
	\equiv\begin{cases}
		\bar{D}_{a}^{(q)} \,(a=S,P,T)\\
		i\bar{D}_{a}^{(q)},(a=V,A)\\
	\end{cases} .
\end{equation*}

On the other hand, in the  epsilon parametrization the effective Lagrangian is expressed as 
\begin{equation}\label{eq:GNI-operators2}
{\cal L} \supset  \frac{G_{F}}{\sqrt{2}}  ~  \sum^{10}_{j=1} \tilde{\varepsilon}^{~ q,   ~ j}_{\alpha \beta \gamma \delta}(  \overline{\nu}_{\alpha} \mathcal{O}_j \nu_{\beta} )  ~ (  \overline{q}_{\gamma} \mathcal{O}^{\prime}_j q_{\delta} ) \;,
\end{equation}
 where operators $ \mathcal{O} $, and $ \mathcal{O}^{\prime} $ take 10 different forms  as given by Table~\ref{tab:GNI-operators}.
 %%%%%%%%%%%%%%%%%%%%%%%%%%%%%%%%%%%%%%%%%%%%%%%%%%%%%
\begin{table}
	\begin{tabular}{!{\hspace{2em}} c !{\hspace{2em}}  | !{\hspace{2em}} c  !{\hspace{2em}} | !{\hspace{2em}}  c !{\hspace{2em}} | !{\hspace{2em}}  c !{\hspace{2em}}}
				$j$ & $  \widetilde{\varepsilon}^{j}$ & $  \mathcal{O}_j $  &  $ \mathcal{O}^{\prime}_j  $\\
		\hline \hline
        1& $ \varepsilon_L $ & 		$\gamma_\mu  \projl $ & $\gamma^\mu \projl $  \\
        2&  $\widetilde\varepsilon_L $ & $ \gamma_\mu \projr $& $\gamma^\mu \projl   $\\
       3&  $\varepsilon_R $ &	$	\gamma_\mu \projl  $ & $  \gamma^\mu \projr  $ \\
       4&  $\widetilde\varepsilon_R $ & $ \gamma_\mu \projr $& $ \gamma^\mu \projr $\\
      5&   $\varepsilon_S $ &	$	\projl $ & 1\\
      6&   $ \widetilde\varepsilon_S $ & $ \projr $ & 1  \\
     7&   $-\varepsilon_P  $&	$	\projl   $ & $ \gamma^5 $\\
      8&  $-\widetilde\varepsilon_P$ & $ \projr $ & $\gamma^5 $\\
      9&   $\varepsilon_T  $ &		$\sigma_{\mu\nu} \projl $ &$ \sigma^{\mu\nu} \projl  $\\
        10&  $ \widetilde\varepsilon_T $ &  $ \sigma_{\mu\nu} \projr $ &$ \sigma^{\mu\nu} \projr$ \\
	\end{tabular}
	\caption{List of GNI operators $ \mathcal{O} $, and $ \mathcal{O}^{\prime} $ as defined by Eq.  \eqref{eq:GNI-operators2}. }
\label{tab:GNI-operators}
\end{table}

Therefore,  after using different $\Gamma^{a} $ in Eq. (\ref{eq:Lagrangian-GNI-1}) and expanding, one can relate these two parametrisations as~\cite{Bischer:2019ttk}
\[\label{eq:Eps-CD-Relation}
\begin{split}
\varepsilon^L &= \frac14\left(C^V-D^V+C^A-D^A\right),\\
\varepsilon^R &= \frac14\left(C^V+D^V-C^A-D^A\right),\\
\varepsilon^S&=\frac12\left(C^S+iD^P\right),\\
-\varepsilon^P&=\frac12\left(C^P+iD^S\right),\\
\varepsilon^T&=\frac14\left(C^T-iD^T\right),
\end{split}
\qquad
\begin{split}
\widetilde\varepsilon^L &= \frac14\left(C^V-D^V-C^A+D^A\right),\\
\widetilde\varepsilon^R &= \frac14\left(C^V+D^V+C^A+D^A\right),\\
\widetilde\varepsilon^S&=\frac12\left(C^S-iD^P\right),\\
-\widetilde\varepsilon^P&=\frac12\left(-C^P+iD^S\right),\\
\widetilde\varepsilon^T&=\frac14\left(C^T+iD^T\right)\;.
\end{split}
\]

\section{$\chi^2$ analyses for COHERENT-LAr data }
\label{app:LAr_analyses}

%%%%%%%%%%%%%%%%%%%%%%%%%%%%%%%%%%%%%%%%%%%%%%%%%%%%%
\begin{figure}[t]
	\includegraphics[scale=0.45]{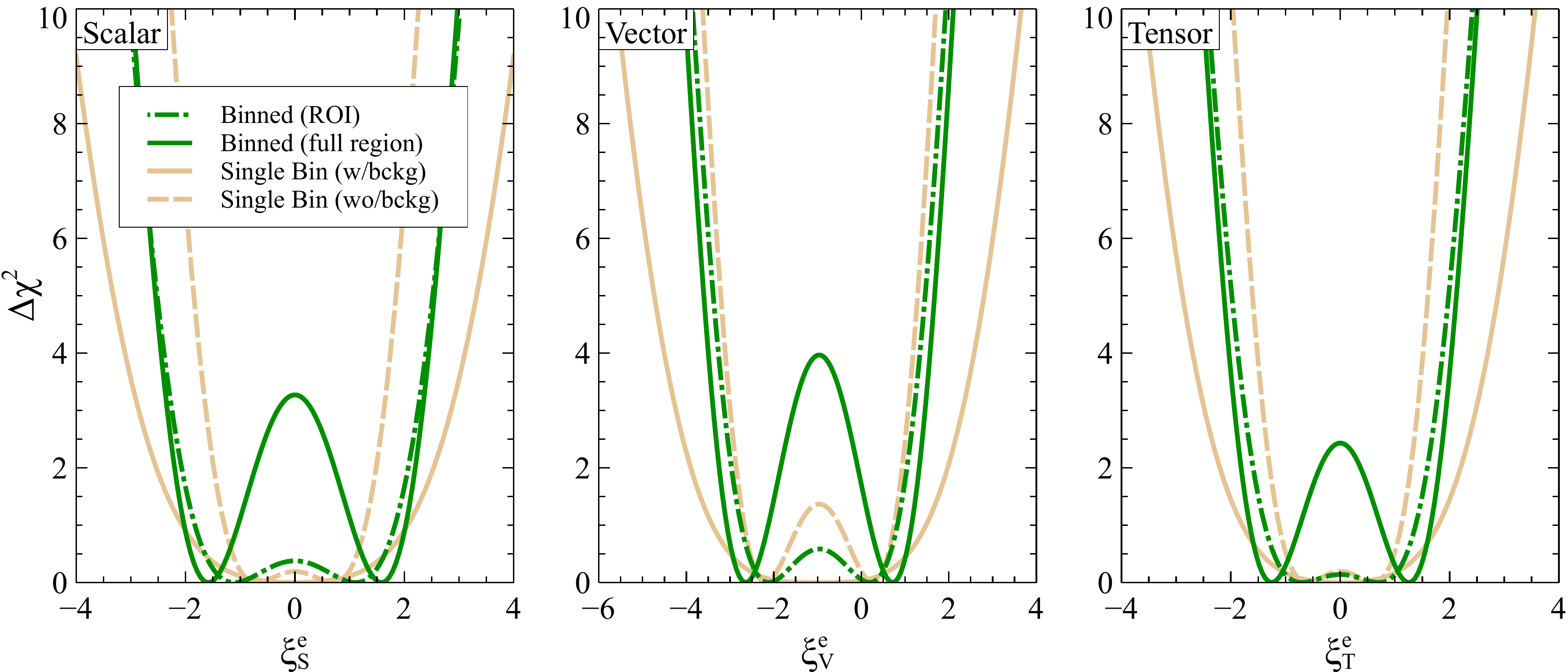}
		\includegraphics[scale=0.45]{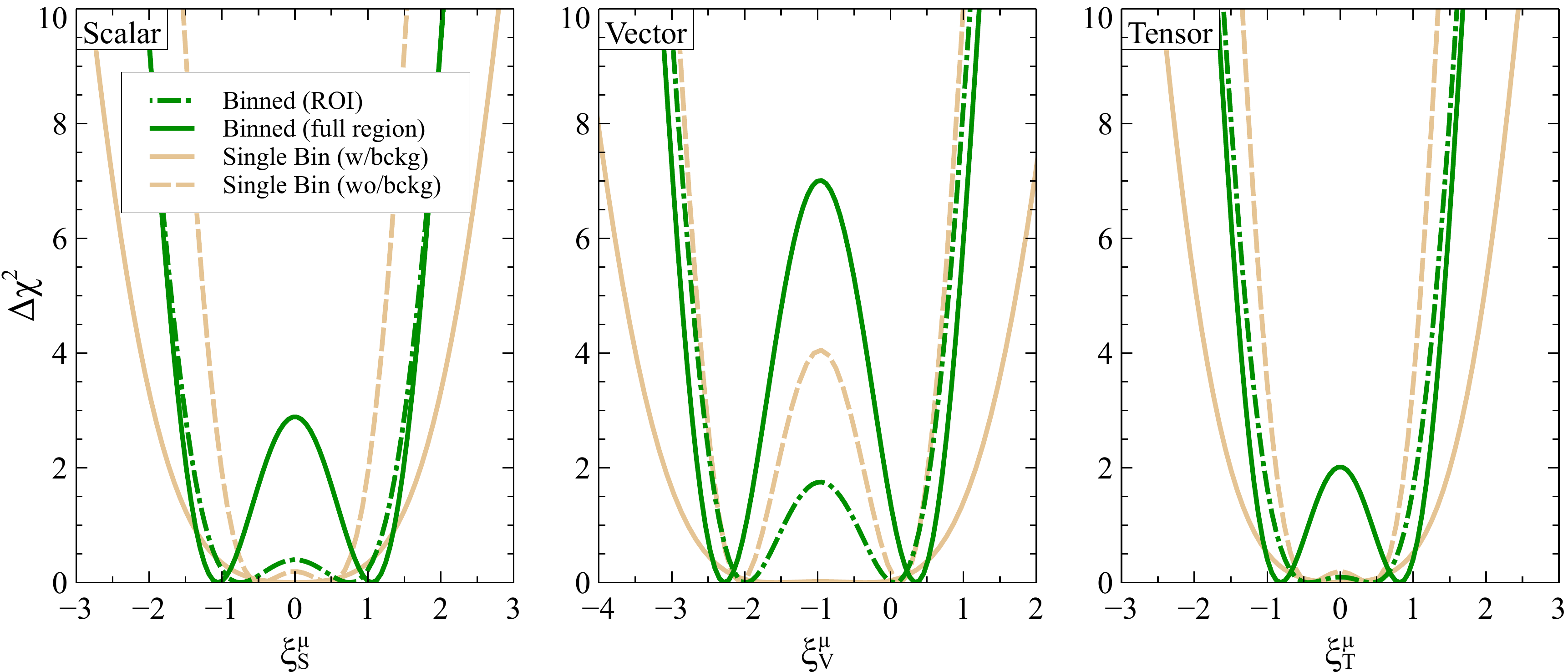}
	\caption{\footnotesize $ \Delta\chi^2 $ for different GNI parameters, where top (bottom) row corresponds to electron (muon) flavor.}
	\label{fig:LAr_methods}
\end{figure}
%%%%%%%%%%%%%%%%%%%%%%%%%%%%%%%%%%%%%%%%%%%%%%%%%%%%%

Since the COHERENT collaboration announced the first measurement of CE$\nu$NS with their CENNS-10 detector, several studies have been performed in order to extract information from this measurement, like the value of the weak mixing angle, limits for some BSM scenarios such as new light vector and scalar mediators, sterile neutrino transition magnetic moments, or even Weakly Interacting Massive Particle (WIMP) discovery limits, just to mention a few~\cite{Miranda:2020tif, Flores:2020lji,
Cadeddu:2020lky,Cadeddu:2020nbr,Banerjee:2021laz, Miranda:2021kre, AristizabalSierra:2021kht}. This can be achieved by means of a $\chi^2$ analysis of the experimental data provided by the COHERENT collaboration. However, there is not a definite recipe for doing such an analysis.

In this appendix, we present different approaches followed in some of the studies mentioned above. 
One approach is to use only the total number of events, which we refer to as a \emph{Single bin} analysis. This method can be applied either including or not the background from the SNS. The other more common approach is the \emph{binned} analysis, in which the available spectral information is used. The COHERENT collaboration provided measurements up to 120 keV$_{ee}$, and some studies have been made using this full spectrum. Nevertheless, we have a significant number of CE$\nu$NS events up to $\sim30$ keV$_{ee}$. The measurements reported above this number correspond mainly to the background. For this reason, in this work, we considered a binned analysis in the region of interest (ROI), which corresponds to the first three energy bins.

In Fig.~\ref{fig:LAr_methods}  we show the limits for the electron (upper panel) and muon (lower panel) GNI parameters, obtained with the four approaches described above: binned (ROI and full region) and single-bin (with and without background) analyses. Regarding both binned methods, we can see that the  two-fold degeneracy is stronger in the full region binned case, whereas it is almost lost at $90\%$ C. L. for the  binned ROI analysis. The reason is that the experimental data in the highest energy region (where CE$\nu$NS is not allowed) fits very well the expected background, therefore resulting in stronger constraints. As for the single-bin methods, the difference between including the background or no, is that the statistical error increases in the former case, hence resulting in weaker constraints.

\bibliography{references}

\end{document}